\documentstyle[pre,aps]{revtex}
\draft
\begin{document}

\title{Disclination loop behavior near the nematic--isotropic transition}
\author{N.~V. Priezjev and Robert A. Pelcovits}
\address{Department of Physics, Brown University, Providence, RI 02912}
\date{\today}
\maketitle

\begin{abstract}
 
We investigate the behavior of disclination loops in the vicinity 
of the first order nematic--isotropic transition in the Lebwohl--Lasher and related models.
We find that two independent measures of the transition temperature, the free energy and the distribution of disclination line segments, give essentially identical values. We also calculate the distribution 
function $D(p)$ of disclination loops of perimeter $p$ and fit it to a quasiexponential form. Below the transition, $D(p)$ falls off exponentially, while in the neighborhood of the transition it decays with a power law exponent approximately equal to 2.5, consistent with a ``blowout" of loops at the transition. In a modified Lebwohl--Lasher model with a strongly first--order transition we are able to measure a jump in the disclination line tension at the transition, which is too small to be measured in the Lebwohl--Lasher model. We also measure the monopole charge of the disclination loops and find that in both the original and modified Lebwohl--Lasher models, there are large loops which carry monopole charge, while smaller isolated loops do not. Overall the nature of the topological defects in both models is very similar.
\end{abstract}

\section{Introduction}
\label{sec:Introduction}

In many physical systems, particularly two--dimensional with continuous
order parameter symmetry, topological defects can play an essential role at
the phase transition
between the ordered (or quasi--ordered) and disordered phases. Using simple
physical arguments Kosterlitz and Thouless \cite{Kosterlitz:73} pointed out
that phase transitions in two--dimensional superfluid $^4\rm{He}$,
crystalline solids and XY magnets would occur via the unbinding of point
topological defects. Subsequently, Kosterlitz \cite{Kosterlitz} developed a renormalization
group theory for the two--dimensional XY model which provided quantitative
predictions for this defect--mediated critical behavior. Renormalization
group theories for the defect--mediated melting of crystalline solids in
two--dimensions were developed by Nelson and Halperin \cite{Nelson:79} and
Young \cite{Young}.

The theoretical picture for defect--mediated phase transitions in three
dimensions is less clear. It is expected on the basis of the Villain representation \cite{Villain} that the transition in the three--dimensional XY model is mediated by vortex loops \cite{Savit}. 
In this scenario there is a finite length scale at
low temperatures associated with the typical size of vortex loops which
disorder the system on smaller length scales. At the phase transition, loops
can exist on all length scales, i.e. there is ``vortex--loop blowout" \cite{Onsager,Feynman} and the system enters the disordered phase. Scaling and renormalization group theories have been developed
\cite{Shenoy-Williams} which provide quantitative predictions for the
critical behavior, though these theories
are not as well established as the corresponding Kosterlitz--Thouless theory
in two dimensions. Monte Carlo simulations \cite{Kohring,Sudbo:98} have yielded further support for this vortex loop picture of the phase transition. In the three--dimensional Heisenberg model there is also numerical evidence \cite{Lau} for a phase transition mediated by point topological defects (monopoles), though this evidence has been questioned by other authors \cite{Janke}.

The role of topological defects at the nematic--isotropic phase transition poses intriguing questions.  The defect structure in
nematics is particularly rich and while sharing some similarities with the
defect structures of the XY and Heisenberg models, there are significant
differences due to the nature of the order parameter space. The local
director $\bf\hat n$ of a nematic liquid crystal is defined as the
average direction of alignment of a group of molecules. However, unlike the
case of ferromagnets, the directions $\bf\hat n$ and $-\bf\hat n$ are
equivalent. Thus, the order parameter space for nematics is  $P_{2}$, the
unit sphere with antipodal points identified. The stable topological defects
\cite{deGennes:93} include monopoles and disclination lines. The
monopoles are similar to the point defects of the Heisenberg model, though
in the latter case, positive and negative topological charges of the same
absolute value are distinct, whereas they are equivalent in the nematic
\cite{Volovik}.
While defect lines in the XY model can have any integer value with either
positive or negative sign, one value of topological charge, $+1/2$,
characterizes the entire class of stable line defects in nematics
\cite{Volovik,Mermin}. All other half--integer valued lines, whether
positive or negative, can be continuously deformed to a line with charge
$+1/2$, and integer--valued lines can ``escape in the third dimension''
\cite{Meyer}. As in the XY model the disclination lines form closed loops or
terminate on the surface of the sample because of the prohibitive energy
cost of a free line end. However, whereas the XY model loops carry no net monopole charge, nematic disclination loops {\it can} (though all need not) carry monopole charge \cite{loops}.

Not only is the classification of defects different in the nematic compared
with the XY and Heisenberg models, but also the nature of the phase
transition, first order in the former case and continuous in the latter.
Lammert et al.~\cite{Toner:95} have argued that disclination lines are
responsible for the first order nature of the nematic--isotropic (NI) transition. They developed
and studied a lattice model of liquid crystals which allows for the suppression of the line defects while maintaining the presence of monopoles as in the Heisenberg model. As the defect lines are made more costly Lammert et al. found that the
transition between the ordered nematic phase (characterized by a nonzero
disclination line tension) and the isotropic phase (characterized by zero
line tension) becomes more weakly first order. At sufficiently large values
of the core energy the transition splits into a pair of continuous
transitions, and a new phase appears with no long--range nematic order but
nonzero disclination line tension. In the extreme limit where the disclination lines are completely suppressed, their model reduces to the Heisenberg model with the expected continuous phase transition.

In this paper we report on a numerical study of topological defect
behavior near the NI transition in the Lebwohl--Lasher (LL)
lattice model \cite{Lasher:72} of liquid crystals which exhibits a weakly first--order phase transition (a weak first--order transition is characteristic of real experimental systems as well). We also consider the defect behavior in a modified LL model which exhibits a strongly first--order transition. We find that two independent measures of the NI transition temperature,  one the free energy and the other the distribution of disclination line segments, give almost identical values. By measuring the distribution of disclination loops as a function of their perimeter, we find evidence for the ``blowout" of disclination loops at the transition, similar to the three--dimensional XY model. The disclination line tension in the modified LL model drops discontinuously to zero at the transition; in the case of the LL model, the transition is too weakly first--order to detect a similar discontinuity. We also measure the monopole charge of the loops and find in both the LL model and its modification that large loops and small loops adjacent to large loops have nonzero monopole charge, while small isolated loops do not. There appear to be no significant differences between the two models in the nature of the topological defects present.

In Sec.~\ref{sec:Simulations and Results} we provide the details of our simulations and results, followed in Sec.~\ref{sec:Conclusions} by our conclusions.

\section{Simulations and Results}
\label{sec:Simulations and Results}

We performed Monte Carlo simulations on the Lebwohl--Lasher model, a
lattice model of rotors with an orientational order--disorder transition.
While it neglects the coupling between the orientational and translational
degrees of freedom present in a real nematic liquid crystal,
it is generally believed that this coupling does not play
a significant role at the NI transition.
With the absence of translational degrees of freedom the LL model is
particularly well--suited for large--scale simulations of the transition.
The model is defined by the Hamiltonian:

\begin{equation}
\label{LL}
{\cal {H}_{LL}}= -J \sum _{<ij>}P_2({\mathbf \sigma}_{i} \cdot  {\mathbf
\sigma}_{j})=
 -J \sum _{<ij>}\biggl\lbrace {3 \over 2}
({\mathbf \sigma}_{i} \cdot  {\mathbf \sigma}_{j})^2 -{1\over
2}\biggr\rbrace
\end{equation}
where the sum is over all nearest-neighbors rotors situated on a cubic lattice. The long axes of the rotors are
specified by the unit vectors ${\mathbf \sigma}_i$), $P_2$ is the 
second--order Legendre polynomial, and $J$ is a coupling parameter. 
The LL model has been intensively investigated using Monte Carlo 
techniques since its introduction
\cite{Luckhurst,Fabbri,Shrock,Biscarini,Zhang:93,Boschi,Zannoni}.
The most complete numerical analysis of the NI transition in the LL model
using the conventional single spin flip Metropolis algorithm was carried out
by
Zhang et al.~\cite{Zhang:93} on systems up to a size of $28^3$. However, the single
spin flip algorithm is inefficient in the critical region and during the course of a simulation the system
becomes trapped in one of the local minima corresponding to either the
ordered or disordered phases. This difficulty can be overcome by using a cluster algorithm which is most 
efficient in the critical region and the system samples both local minima 
effectively.. The first such algorithm for nematic liquid crystals was introduced by Kunz and Zumbach \cite{Kunz} to study the two--dimensional LL model. Their algorithm is a modification of the Wolff algorithm \cite{Wolff:89} for ferromagnetic systems, and
greatly reduces critical slowing down. In an earlier publication \cite{Priezjev:01}
we used the Kunz--Zumbach algorithm to carry out a finite--size scaling analysis 
of the NI transition in the three--dimensional LL model with systems sizes up to $70^3$ using the Ferrenberg--Swendsen 
reweighting technique \cite{Ferrenberg:89}. 

Here we use the cluster algorithm to study the behavior of disclination
lines and monopoles in the critical region. Following Ref. \cite{Toner:95}
we introduce a disclination line segment counting operator,

\begin{equation}
\label{D}
D_{ijkl} \equiv \frac{1}{2}\Bigl\lbrack{1-{\rm sgn}\{(\mathbf \sigma}_{i} \cdot
{\mathbf
\sigma}_{j})({\mathbf \sigma}_{j} \cdot {\mathbf \sigma}_{k})({\mathbf
\sigma}_{k} \cdot {\mathbf \sigma}_{l})({\mathbf \sigma}_{l} \cdot {\mathbf
\sigma}_{i}) \}\Bigr \rbrack
\end{equation}
which is unity if a disclination line segment pierces the lattice square
defined
by the four rotors ${\mathbf \sigma}_i,{\mathbf \sigma}_j,{\mathbf
\sigma}_k$ and ${\mathbf \sigma}_l$. This method of locating disclination
segments is mathematically equivalent to the method of Zapotocky et
al.~\cite{Zapotocky:95}. A disclination line segment can be considered as a
bond on a cubic lattice dual to the original lattice of the rotors, and only
an even number of bonds meet at a dual lattice site. Connecting the bonds to
form disclination loops cannot be done in a unique way when four or six
bonds meet at a site. 
To deal with this ambiguous case we followed the approach of
Ref.~\cite{Sudbo:98} and chose a random pairing of the bonds. We thus
traced the path of each disclination line through the system until the path
crossed itself and formed a loop.
The bonds of the loop were then eliminated from the dual lattice to avoid
double counting when additional loops were traced. 

We considered several measures of the nature of the disclination segments and loops, including the number of segments in the nematic and isotropic phases at coexistence, the distribution of loops as a function of their perimeter, and the monopole charge of the loops.

We simulated a system of size $70^3$ (with periodic boundary conditions)
for $5 \times 10^6$ Monte Carlo steps (MCS) where one MCS
corresponds to one cluster formation attempt and update of the spins
comprising the cluster. Every 200 MCS we measured the total number of disclination line segments in the system and stored our data in a histogram. The logarithm of this histogram is shown in Fig.~\ref{bonddensity} at the temperature $T=1.1226$ (temperatures measured in units of $J/k_{B}$)
where the two wells have equal depth. The right--hand well corresponds to the isotropic phase (which we confirm by monitoring the nematic order parameter) and the left--hand well with fewer disclination line segments corresponds to the nematic phase. Varying the temperature by as little as 0.0001 yields wells of unequal depths. In our earlier work \cite{Priezjev:01} we used the cluster algorithm to compute a histogram of the free energy (as a function of $E={\cal {H}_{LL}}/N$, the energy per site). We found a double well structure for the free energy with equal well depths occuring at the temperature $T=1.1226$, which we then identified as $T_{NI}$. Thus, the appearance of equal well depths in the histogram for the density of disclination line segments at the {\it same} temperature (to within our numerical accuracy) suggests that disclinations play a crucial role at the NI transition.

To further assess the role played by disclination loops at the transition we followed the approach used in Ref.~\cite{Sudbo:98} to study defect behavior at the three--dimensional XY transition, and calculated the perimeter distribution function $D(p)$ (the average number of loops with perimeter $p$). In Refs. \cite{Sudbo:98} $D(p)$ was fit to
the following form:

\begin{equation}
D(p)=A  p^{-\alpha}  \exp(- \epsilon(T) p / k_{B} T),
\label{Dp}
\end{equation}
where $\epsilon(T)$ is the effective vortex line tension which is non zero
at low temperatures. Thus, in the low temperature ordered phase vortex loops
with large $p$ are
exponentially suppressed, and the length scale $L_{0}$ governing the typical
perimeter size of the vortex loops is given by $L_{0}=k_{B}T / \epsilon(T)$.
At the critical temperature $T_c$ of the XY model, $\epsilon(T)$ vanishes continuously and the distribution $D(p)$ has a power--law form. Consequently there will
be a finite probability
of having vortex loops which traverse the entire system and destroy the
long--range order. 

We computed $D(p)$ at the NI transition temperature $T_{NI} = 1.1226$ (identified by the two methods described above)
and at two slightly lower temperatures.
The results are shown in Fig.~{\ref{figloops}. We simulated the system
for $5 \times 10^6$ MCS and computed $D(p)$ every $200$ MCS to make sure that the successive 
configurations for bonds distribution are completely updated. 
Fitting our data to the form Eq.~(\ref{D}) yields $\alpha=2.50 \pm 0.05$. For
noninteracting loops (i.e. random walks) we expect $\alpha=2.5$ exactly, whereas $\alpha>2.5$
and $\alpha<2.5$ for repulsive and attractive loop interactions respectively
\cite{Antunes:98}. Unfortunately, we cannot distinguish among these three
possibilities in our data.

While the relative depths of the wells appearing in the histogram Fig.~\ref{bonddensity} are sensitive to temperature variations as small as 0.0001 about $T_{NI}$, the distribution $D(p)$ is less sensitive. A plot of $D(p)$ at $T=1.1225$, e.g., would be qualitatively similar to the appearance of $D(p)$ at $T_{NI}$, the upper curve in Fig.~\ref{figloops}.  However,  
deeper in the nematic phase at $T=1.120$ (where the isotropic free energy well has disappeared) and $T=1.10$, the behavior of $D(p)$ is different. Here one
can clearly see an exponential decay at large values of $p$ and the data can be fit
to the form given in Eq.~(\ref{Dp}) with line tensions $0.003$ and 0.026 at $T=1.120$ and $T=1.10$ respectively (these values were computed setting $\alpha=2.5$).

The behavior of $D(p)$ shown in Fig.~\ref{figloops} is consistent with a ``blowout" of disclination loops at the NI transition, similar to the behavior found in the three--dimensional XY model \cite{Sudbo:98}. However, as indicated in the previous paragraph, $D(p)$ provides a less sensitive measure of the transition temperature compared to the histogram of the disclination line segments. The source of this drawback is the appearance of the bump at large perimeters in $D(p)$ in the neighborhood of $T_{NI}$ which is a
finite--size effect, arising from the nonzero probability
of forming loops which wrap completely around the lattice due to the periodic boundary conditions. At high enough
temperatures (in particular near the transition and above)
there will be a sufficient number of disclination line segments present to
form such loops.
The value of $\alpha$ for these loops is predicted to be unity for
noninteracting loops in three dimensions \cite{Austin:94}. 
The crossover in $D(p)$ from infinite
system behavior (with $\alpha \simeq 2.5$) to the finite system behavior is
expected to occur at a critical perimeter
$p_{c}(L) \approx 1.5 L^2/ \pi  $ for a system of size $L$ \cite{Austin:94}. 
Our results are in very good agreement with this scenario 
as shown in Fig.~\ref{figloops}.
The truncation of $D(p)$ at very large perimeters occurs
because there are not enough disclination line segments 
(since we eliminated the previously marked ones)
to construct loops with arbitrarily large perimeters.
The deviation from the expected $\alpha=2.5$ behavior for small $p$ is due
to the presence of the underlying lattice structure of the LL model.

We attempted to measure the jump in the effective line tension
$\epsilon$ at $T_{NI}$ in the LL model which is expected to occur because of
the first order nature of the NI transition, in contrast to the continuous
vanishing of $\epsilon$ observed in the XY model \cite{Sudbo:98}.
Naively one should compute $D(p)$ for temperatures
in the vicinity of $T_{NI}$ and extract the line tension $\epsilon(T)$ using
Eq.~(\ref{Dp}) to fit the data.
However, in practice this procedure is very difficult to carry out because
the NI transition is weakly first order in the LL model and
the jump in $\epsilon$ must be determined from the behavior of loops with
very large perimeters.
But loops with large perimeters wrap around the lattice as we described in
the previous paragraph and thus appear in the ``bump" region of 
Fig.~\ref{figloops} which cannot be fit with Eq.~\ref{Dp}.

Another possibility would be to set the temperature to a value 
close to $T_{NI}$, calculate the nematic order parameter and then 
compute $D(p)$ separately in the isotropic and nematic phases. 
In principle the difference between the two distributions
should give the jump in $\epsilon$ at the given temperature. However, again
due to the very weak first order nature of the transition we found that each
of these distributions was qualitatively similar to the critical
distribution shown in Fig.~\ref{figloops}. Thus at least for the systems
sizes we have been able to study (less than or equal to $70^3$), we have found
it impossible to accurately measure the expected jump in the disclination 
line tension in the LL model.

To check our supposition that the disclination line tension should have a
discontinuous jump at the first order NI transition, we considered a
modified LL model, including a fourth--order Legendre polynomial $P_4$, which has been shown to make the NI transition more strongly first
order \cite{Zhang:93,Zannoni}. The modified Hamiltonian is given by:

\begin{equation}
\label{modLL}
{\cal {H}^\prime} = -J \sum _{<ij>}P_2({\mathbf \sigma}_{i} \cdot
{\mathbf \sigma}_{j}) - J^\prime \sum _{<ij>}P_4({\mathbf \sigma}_{i}
\cdot  {\mathbf \sigma}_{j})
\end{equation}
This modified LL model was studied in Ref.~\cite{Zhang:93} for a system of size $24^3$ with $J^\prime/
J = 0.1$. We checked that the cluster algorithm produces free energy
plots similar to those obtained in the latter reference where the single
flip Monte Carlo algorithm was used.

We were able to measure the jump in the line tension for a system of 
size $50^3$ with $J^\prime / J = 0.3$ as shown in
Fig.~\ref{figeps4}, finding a jump of approximately  $0.0024$
at the transition temperature $T=1.2475$.
Here we calculated $D(p)$ separately in the isotropic and nematic phases, with a production run of $10^7$ MCS. 
We have been unable to carry out a finite--size scaling analysis of the jump
because one has to choose the ratio
$J^\prime / J$ large enough so that $D(p)$ exhibits behavior
consistent with the form of Eq.~\ref{Dp} at large $p$ in nematic phase; 
in particular with no ``bump" as in Fig.~\ref{figloops}. 
However, this choice of $J^\prime/ J$ makes the transition 
more strongly first order and large systems can hardly overcome the resulting 
large free energy barrier.
We also measured a histogram of the density of disclination line segments for the modified model and found behavior similar to that found in the LL model (see Fig.~\ref{bonddensity}), namely equal well depths for this quantity at the {\it same} temperature where the free energy also exhibits two wells of equal depth.

As we discussed in Sec.~\ref{sec:Introduction}, topological defects in nematics include not only disclination loops but monopoles as well, and furthermore the disclination loops can potentially carry monopole charge.
To locate monopoles we used the prescription introduced by Berg and L\"uscher \cite{Berg} for the two--dimensional Heisenberg model which was subsequently extended by Lau and Dasgupta \cite{Lau} to three dimensions. Each of the
six faces of a lattice cube is divided into two equal area triangles by the face diagonal. The rotors ${\mathbf \sigma}_{1},{\mathbf \sigma}_{2}$ and ${\mathbf \sigma}_{3}$ at the three corners of each of the triangles are mapped to points on the order parameter sphere, forming spherical triangles. The area of each of the twelve spherical triangles formed by this mapping is then computed, with a sign given by $\rm{sgn}({\mathbf \sigma}_{1}\cdot({\mathbf \sigma}_{2}\times{\mathbf \sigma}_{3}))$. The rotors on each triangle are numbered so that the circuit $1 \rightarrow 2 \rightarrow 3 \rightarrow 1$ corresponds to a counterclockwise rotation along the outward normal to the surface of the triangle. In performing this mapping we have assigned heads to the rotors such that the distance between the heads on the order parameter sphere is minimized, i.e., we use the ``geodesic rule" \cite{vach,rudaz,hind-brand} to effectively minimize the energy. For lattice cubes which are not pierced by disclination lines (and it is these cubes that we examine for monopole charge), the heads of all eight rotors at the corners of a cube can be simultaneously chosen to obey the geodesic rule without frustration. Thus, the angle between any pair of rotors at the corners of any of the twelve triangles will no greater than $90^\circ$. Finally, the monopole charge enclosed by the cube is given by the sum of the twelve signed areas of the spherical triangles. 

Using this algorithm we examined all lattice cubes which are not pierced by disclination lines and found {\it no} monopoles, neither in the original LL model, Eq.~(\ref{LL}), nor the modified model, Eq.~(\ref{modLL}). We searched for monopoles in the neighborhood of $T_{NI}$, deep in the nematic phase and at very high temperatures; in all cases no monopoles were located. This null result is not surprising given the topological arguments advanced by Hindmarsh \cite{hind} which yield a very low probability (of order $10^{-8}$, compared with 1/8 for the Heisenberg model) for the appearance of point monopoles in a nematic.

One way to measure the monopole charge of a disclination loop would be to apply the above algorithm to the surface of a group of lattice cubes which completely enclose a disclination loop. We have carried out this procedure for isolated loops of perimeter $p=4$; none of these loops were found to carry monopole charge. It is difficult to carry out this procedure for larger loops, especially when two or more loops are entangled. In particular, when loops are entangled it is impossible to impose the geodesic rule simultaneously on all pairs of rotors. If we surround one loop completely with a set of lattice cubes, frustration will arise where the second loop pierces one of these lattice cubes. Instead we measured the local rotation vector $\mathbf\Omega$ of the four rotors surrounding each of the segments which form a disclination loop, and then summed these vectors along the entire length of the loop. If this sum is nearly zero, then the loop carries a nonzero monopole charge, because the set of rotors surrounding the entire loop will cover essentially the entire order parameter sphere \cite{Lubensky}. A simple example of this topology occurs in the case of a pure wedge loop where $\mathbf\Omega$ is everywhere tangent to the loop \cite{windle}, and summing this vector around the loop yields zero identically; a simple example of a loop with zero monopole charge is a twist loop where $\mathbf\Omega$ is everywhere perpendicular to the plane of the loop.  We measured the local rotation vector $\bf\Omega$ of the four rotors  ${\mathbf \sigma}_{1},{\mathbf \sigma}_{2},{\mathbf \sigma}_{3},{\mathbf \sigma}_{4}$ which lie at the corners of a lattice square pierced by a disclination line, by summing the vector cross product of each neighboring pair of rotors \cite{Hobdell:97}:
\begin{equation}
\label{Omega}
{\mathbf\Omega}= ({\mathbf \sigma}_{1}\times{\mathbf \sigma}_{2})+({\mathbf \sigma}_{2}\times{\mathbf \sigma}_{3})+({\mathbf \sigma}_{3}\times{\mathbf \sigma}_{4})+({\mathbf \sigma}_{4}\times{(-\mathbf \sigma}_{1}))
\end{equation}
In writing this definition of $\mathbf\Omega$ we have chosen the heads of the rotors so that the neighboring pairs ${\mathbf \sigma}_{1},{\mathbf \sigma}_{2};{\mathbf \sigma}_{2},{\mathbf \sigma}_{3}$ and ${\mathbf \sigma}_{3},{\mathbf \sigma}_{4}$ satisfy the ``geodesic rule" on the order parameter sphere. The remaining pair ${\mathbf \sigma}_{4},{\mathbf \sigma}_{1}$ will not satisfy this rule because of the presence of the disclination line segment; thus we reflect ${\mathbf \sigma}_{1}$ in the last term in Eq.~(\ref{Omega}), so that the vector products always involve pairs of rotors that satisfy the geodesic rule.  To assign a unique sense to the circuit $1 \rightarrow 2 \rightarrow 3 \rightarrow 4 \rightarrow 1$, we arbitrarily assign a direction along the length of the disclination loop, and traverse the circuit in a counterclockwise sense along this direction. We note that the reflection of ${\mathbf \sigma}_{1}$ in the last term of Eq.~(\ref{Omega}) guarantees that $\mathbf\Omega$ is independent of which of the four rotors is labeled 1.  

Our results for the vector sum of $\mathbf\Omega$ along each disclination loop are shown in Fig.~\ref{omega} for the LL model at its NI transition temperature $T_{NI}=1.12279$ for a system of size $50^3$. We use a smaller system size because the computation of this vector sum must be done using scalar code, whereas the cluster algorithm used above can be vectorized \cite{Priezjev:01}. Similar results were obtained for the modified LL model.  We note that all of the large loops with perimeters $p \gtrsim 100$ are characterized by net rotation vectors which are nearly zero, suggesting that they carry nonzero monopole charge. We expect on energetic grounds that this charge will be unity rather than higher values. We have checked this supposition for a random sampling of loops finding that the rotation vectors cover a great circle on the order parameter sphere just once. Our data indicates that small isolated loops do not carry monopole charge. Rather the monopole charge is carried by large loops (with perimeters greater than 100) and small loops which touch larger ones.

\section{Conclusions}
\label{sec:Conclusions}
In this paper we have studied the properties of topological defects in two lattice models of the NI transition: the original Lebwohl--Lasher model (which exhibits a weakly first order transition) and a modified model with a more strongly first order transition. We have found evidence for the role played by disclination loops at the NI transition in both models. Namely, a histogram of disclination line segments collected over the course of the MC simulation shows a double well structure, and the wells are of equal depth at the {\it same} temperature where the free energy exhibits similar structure. We also find that the distribution $D(p)$ of disclination loops as a function of their perimeter exhibits power--law behavior at this temperature, consistent with the ``blowout" of loops at the transition. However, $D(p)$ is a less sensitive measure of the transition temperature compared with the disclination segment histogram, due to finite size effects.

We have also searched for point monopoles in these models and measured the monopole charge of the disclination loops. We found no point monopoles, a result that may be reasonable on the basis of topological arguments \cite{hind}.  However, we did find that nearly all of the large disclination loops carry monopole charge, while small isolated loops do not. Of particular interest is the result that the two models we studied, one with a weakly first--order transition and the other with a strongly first--order transition, showed no qualitative differences in their defect characteristics, other than the measurable jump in the disclination line tension in the latter model. In light of the results of Ref.~\cite{Toner:95} we find the similarities in the defect characteristics of the models we studied somewhat surprising. In Ref.~\cite{Toner:95} it was shown that suppressing disclination loops while leaving monopoles yields a more continuous NI transition. One might guess then that moving in the opposite direction to a model like that given in Eq.~(\ref{modLL}) which exhibits a strongly first--order transition one would find fewer monopole--like entities than in the LL model with its weakly first--order transition. While neither model has point monopoles, both appear to have similar densities of disclination loops with monopole charge, suggesting that monopole charge may not influence the strength of the first--order transition. We should also note that while our results suggest that disclination loops ``blowout" at the NI transition in both models we considered, it is not clear from our study whether the transition is in fact defect--driven, or rather that some other mechanism drives the transition and the defects simply respond. Clearly more work on this very intriguing phase transition and the role played by topological defects would be of considerable interest. 
\section*{Acknowledgments}

We thank  A. Sudb\o, J. M. Kosterlitz and J. Toner for numerous helpful discussions.
This work was supported by the National Science Foundation under grant
DMR--9873849.
Computational work in support of this research was performed at Brown
University's Theoretical Physics Computing Facility.

\bibliographystyle{prsty}

\begin{figure}
\caption{The logarithm of the distribution of disclination bond density 
in the Lebwohl--Lasher model, Eq.~(\ref{LL}) for system size $70^3$ at the NI transition 
temperature $T_{NI}=1.1226$. The density is defined as the ratio of the number of disclination bonds to the total number of lattice bonds ($3 \times 70^3$ in the present case). The solid line is a guide to the eye.}
\label{bonddensity}
\end{figure}

\begin{figure}
\caption{Log--log plot of the disclination loop distribution function $D(p)$ (Eq.~(\ref{Dp}))
in the Lebwohl--Lasher model, Eq.~(\ref{LL}) for system size $70^3$
at temperatures: $T_c=1.1226$ (top curve), $T=1.120$ (middle curve)
and $T=1.10$ (bottom curve).  }
\label{figloops}
\end{figure}

\begin{figure}
\caption{Log--log plot of the disclination loop distribution function $D(p)$ for the modified Lebwohl--Lasher model, Eq.~(\ref{modLL}), at its NI transition temperature $T=1.2475$. The system size is $50^3$ and the ratio of the couplings is $J^\prime / J = 0.3$. The top curve (which has been displaced for the sake of clarity) and the bottom curve correspond to the isotropic and nematic wells of the free energy respectively. The jump in the disclination line tension is found to be 0.0024, and the straight portion of the isotropic data can be fit with a power--law of $2.50 \pm 0.01$ (see Eq.~(\ref{Dp})).}
\label{figeps4}
\end{figure}

\begin{figure}
\caption{The distribution of $\mid\Sigma \ \ {\mathbf \Omega}\mid /p$, the magnitude of the vector sum of the rotation vector $\Omega$, Eq.~(\ref{Omega}),
along each disclination loop divided by its perimeter $p$, in the LL model of size $50^3$ at $T_{NI}$. The top curve includes loops of all perimeters, the middle curve includes loops of $p=4$ only, while the bottom curve includes only loops with $p>100$. The rightmost peak appearing in the middle curve corresponds to isolated 
loops. Note that for a perfect wedge line segment piercing a square face of a lattice cube (i.e., a rotor configuration of the form ${\mathbf\sigma}=(\cos{\phi/2},\sin{\phi/2})$, where $\phi$ is the azimuthal angle of the lattice site), $\mid\Sigma \ \ {\mathbf \Omega}\mid$ is given by $2\sqrt{2}\approx 2.8$.}
\label{omega}
\end{figure}

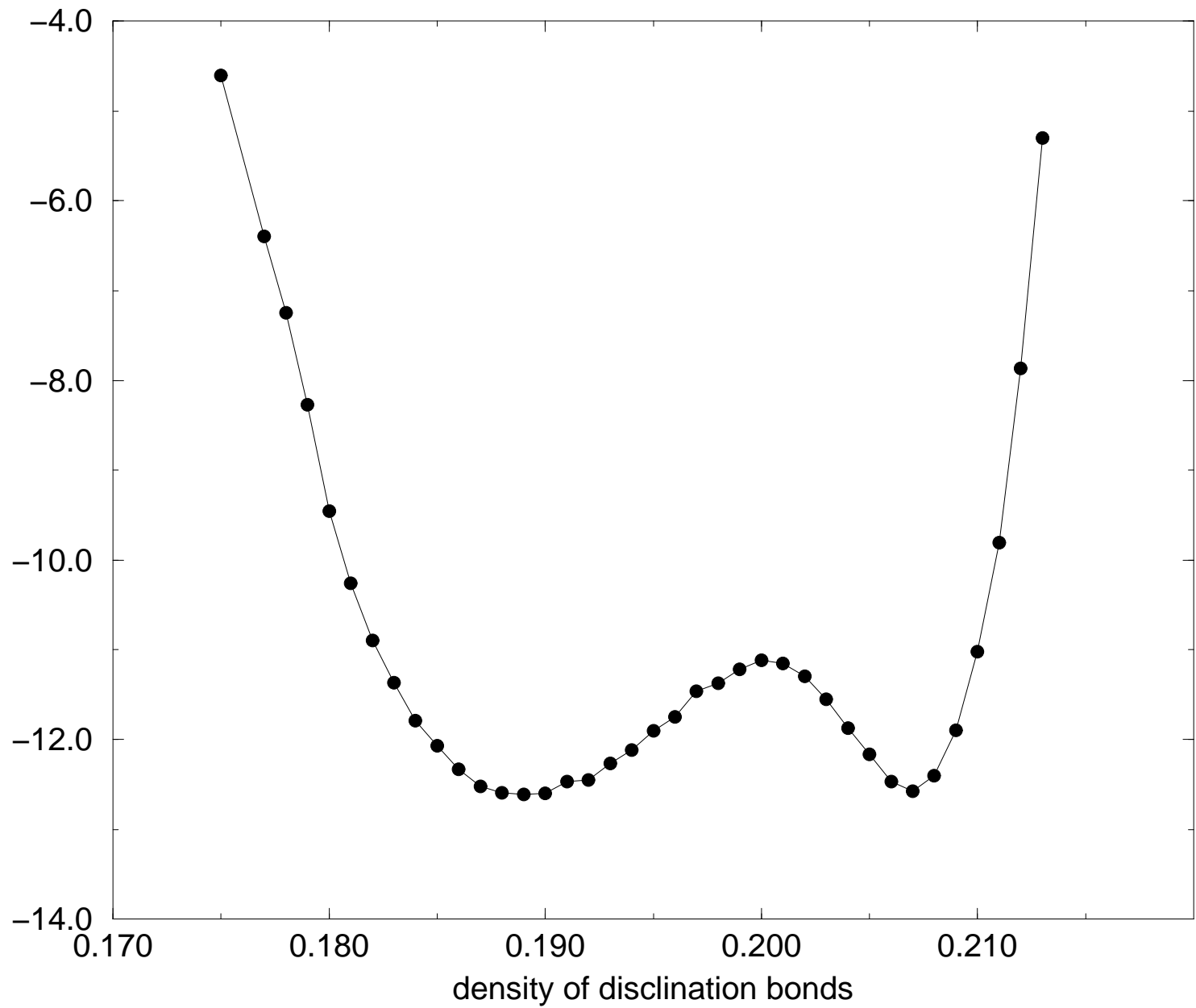
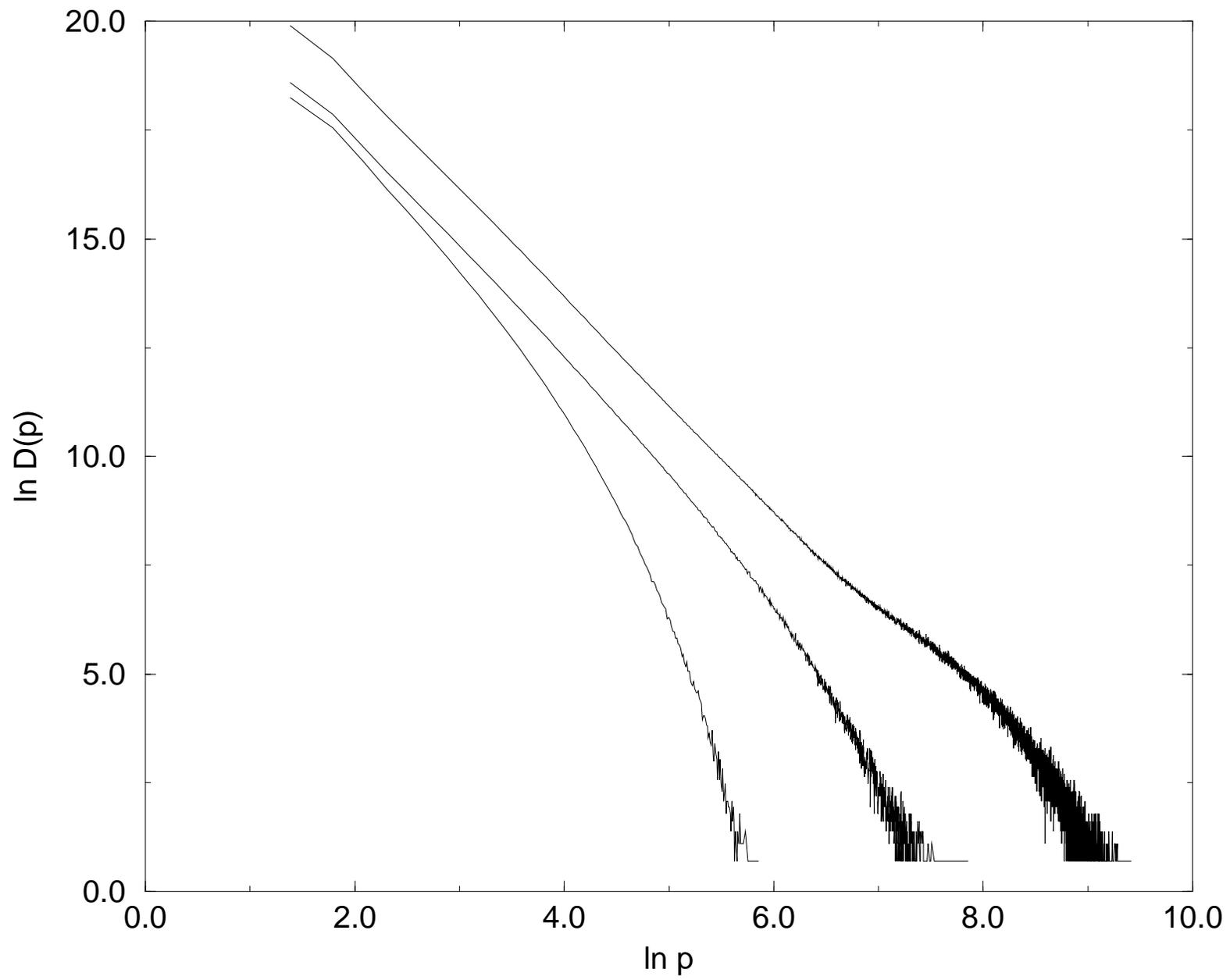
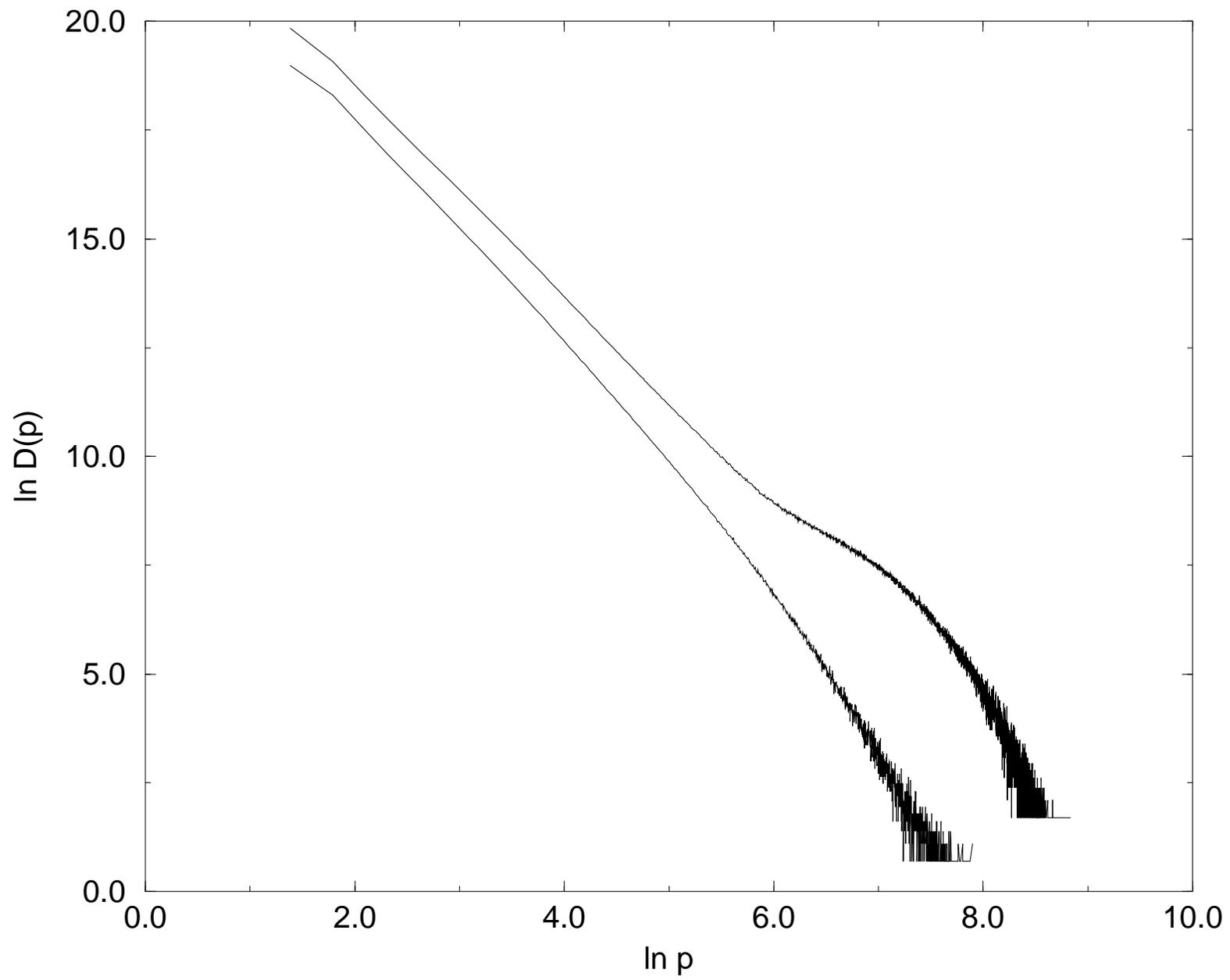
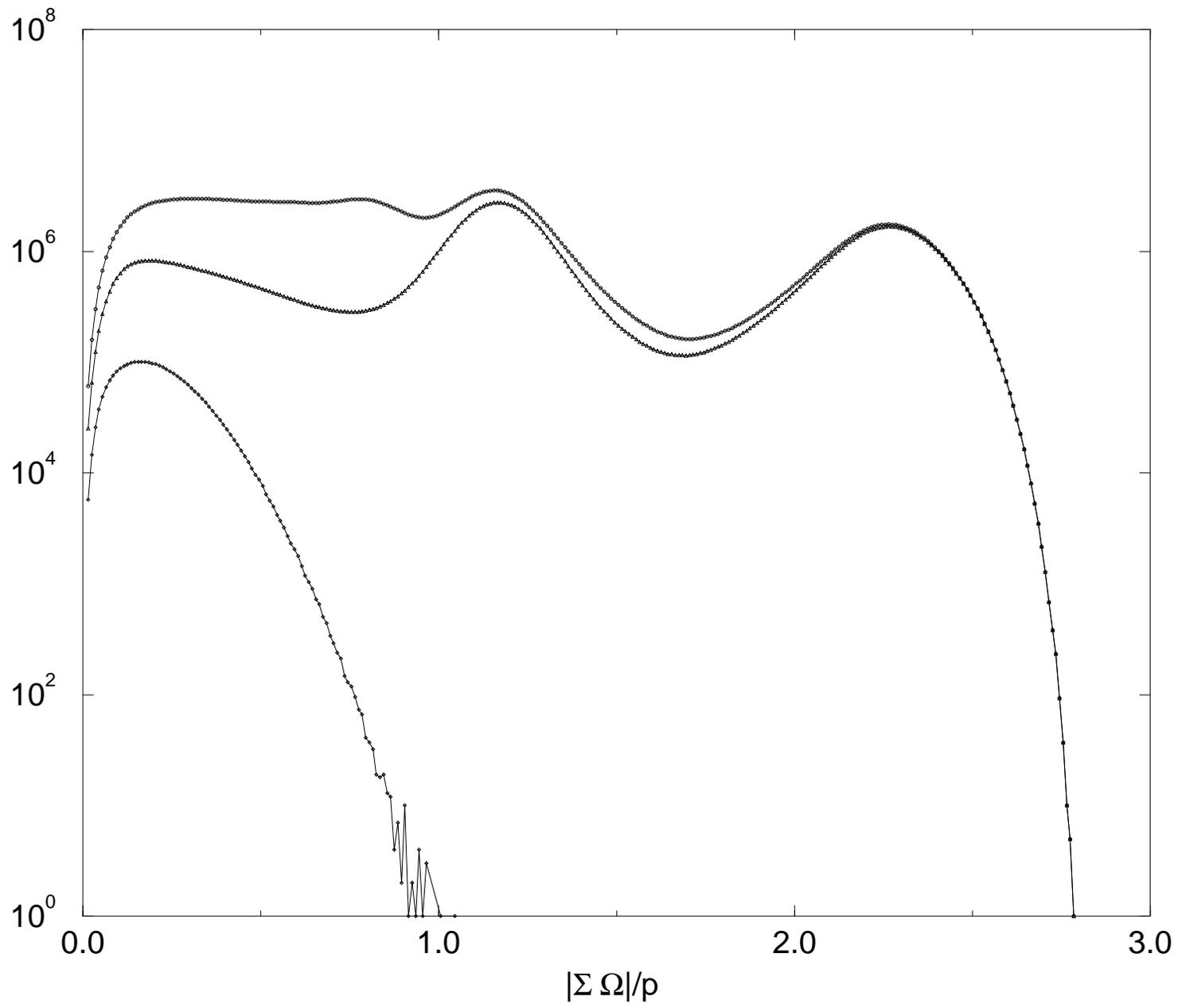
\end{document}